\documentclass[10pt,arxiv,red, onehalfspacing]{lapreprint}
\pdfoutput=1

\usepackage[version=4]{mhchem} 
\usepackage{siunitx}    
\usepackage{graphicx}
\usepackage{pdflscape}  
\usepackage{rotating}   
\usepackage{textgreek}  
\usepackage{gensymb}    
\usepackage[misc]{ifsym} 
\usepackage{listings}   
\usepackage{colortbl}   
\usepackage{tabularx}   
\usepackage{longtable}  
\usepackage{subcaption}
\usepackage{multirow}
\usepackage{snotez}     
\usepackage{csquotes}   

\DeclareSIUnit\Molar{M}

\usepackage[			
	backend=biber,      
    style=authoryear,   
	natbib=true,		
	hyperref=true,	    
	alldates=year,      
]{biblatex}

\title{Did the Musk Takeover Boost Contentious Actors on Twitter?}

\author[1 \Letter]{Christopher Barrie}

\affil[1]{School of Social and Political Sciences, University of Edinburgh}

\metadata[]{\Letter\hspace{.5ex} For correspondence}{\href{mailto:christopher.barrie@ed.ac.uk}{christopher.barrie@ed.ac.uk}}
\metadata[]{Present address}{Chrystal Macmillan Building, 15a George Square, Edinburgh, EH8 9LD, United Kingdom}
\metadata[]{Data availability}{Data availability statement. Replication data will be provided at \url{https://github.com/cjbarrie/muskboost}}

\leadauthor{Christopher Barrie}
\shorttitle{Did the Musk Takeover Boost Contentious Actors}

\begin{document}
\maketitle
\begin{abstract}

Twitter has been accused of a liberal bias in its account verification and content moderation policies. Elon Musk pledged, after his acquisition of the company, to promote free speech on the platform by overhauling verification and moderation policies. These events sparked fears of a rise in influence of contentious actors---notably from the political right. In this article, I use a publicly released list of 138k Twitter accounts that purchased blue check verification during the open window of November 9–November 11, 2022. I retrieve $\sim$4.9m tweets from a sample of politically contentious accounts. I then compare engagement on contentious user posts before and after the Musk acquisition. I find that the period following the Musk acquisition saw a substantive increase in post engagement. There is no additional increase following blue tick verification. I explain the findings with reference to an increase in activity by a newly sympathetic user base.

\end{abstract}
\section{Introduction} \label{intro}

Platform moderation and verification policies are important instruments in the fight against online harm and misinformation. They are also central to ongoing debates about free speech and its limits online. Twitter has been entangled in these debates for two reasons. It has been accused, first, of a liberal bias that operates at the level of the algorithmic ranking system \citep{huszar_algorithmic_2022}. Second, it has been accused of overzealous moderation policies that disproportionately target the political right by removing them from the platform or removing their blue-tick verified status \citep{anyanwu_why_2022}.

Elon Musk himself is one of those who has accused Twitter of harbouring such bias \citep{koenig_elon_2022}. It was on the basis that, following his  acquisition of Twitter in October, 2022, he pledged to overhaul both content moderation and verification procedures. In a tweet posted on November 1, Musk wrote that "Twitter’s current lords \& peasants system for who has or doesn’t have a blue checkmark is bullshit" and vowed to introduce a revamped, paid, "Twitter Blue" verification service to take its place.\footnote{See: \url{https://twitter.com/elonmusk/status/1587498907336118274} or Appendix Figure \ref{fig:musktweetlong} for the full tweet thread.}

Separately, Musk has vowed to overhaul moderation practices on the Twitter platform. A self-described "free speech absolutist," Musk's acquisition of Twitter was welcomed by many quarters critical of what they perceived as Twitter's overzealous platform moderation policies \citep{milmo_how_2022, anyanwu_why_2022}. Elsewhere, Musk has accused Twitter of harbouring left-wing bias---both algorithmic and resulting from enforcement policies \citep{koenig_elon_2022}. In the wake of his acquisition of Twitter, observers detected considerable spikes in hate speech on the platform \citep{network_contagion_research_institute_ncri_io_evidence_2022, anyanwu_why_2022}.

Web platform affordances such as verification provide users with important heuristics for judging the credibility of information \citep{metzger_social_2010, flanagin_role_2007}. Early research found that the blue tick on Twitter had become a shorthand for credibility and legitimacy \citep{morris_tweeting_2012}. More recent experimental interventions suggest that the blue tick no longer serves as a marker of credibility \citep{edgerly_blue_2019, vaidya_does_2019}.

Whether or not verification provides a seal of information credibility for users, verified users continue to play a central role in online information flows. This is most obvious during periods of political unrest when both verified and unverified users have to make additional efforts to determine the credibility of online content \citep{gonzalez-bailon_bots_2021, rauchfleisch_how_2017}. Here, we see that (unverified) bot users are far less visible than verified accounts within the discussion of contentious political events.

Platform moderation policies, on the other hand, such as labelling or removal policies, have been effective in stemming the flow of conspiracy theories online and curbing extremist rhetoric \citep{papakyriakopoulos_spread_2020, ganesh_countering_2020}. Removing contentious users from platforms both reduces conversations about individuals in question and reduces the toxicity of content produced by followers \citep{jhaver_evaluating_2021}.

Against this backdrop, we can develop two hypotheses. Given the stated position of Musk against existing platform moderation policies and the alleged liberal bias of Twitter as a platform, his acquisition of Twitter will attract new or otherwise dormant users to the platform who are sympathetic to his views. In turn, this will boost engagement for contentious actors on Twitter who are aligned with his stated positions. 

\textbf{Hypothesis 1}: \textit{Elon Musk's acquisition of Twitter will boost engagement on contentious user content}.

We also know that blue tick verification may provide a visibility and credibility boost for accounts. Given that any account was able to purchase blue tick verification from November 9--November 11, 2022, we can therefore expect verified users to see an additional boost in engagement from followers after verification. 

\textbf{Hypothesis 2}: \textit{Blue tick verification will boost engagement on contentious user content}.

\section{Data \& Methods} \label{methods}

In order to determine any engagement boost that followed the Musk acquisition, I make use of a public list of accounts that paid for verification during the short window of its initial launch: November 9, 2022--November 11, 2022. This list was compiled and made by public by two software developers, Travis Brown and Casey Ho. The data release was verified and reported on by major international newspapers and civil society organizations at the time.\footnote{See e.g., \url{https://www.washingtonpost.com/technology/2022/11/16/musk-twitter-email-ultimatum-termination/}; \url{https://www.nytimes.com/2022/11/11/technology/twitter-blue-fake-accounts.html}; \url{https://www.splcenter.org/hatewatch/2022/11/16/twitter-blesses-extremists-paid-blue-checks}.}.

These data on the timing of verification are both novel and useful. They are novel because information on the timing of any verification decision is not normally made public through the Twitter API. They are useful as they enable us to determine both the effect of the Musk acquisition and any additional effect of blue tick verification on post engagement by these users. In other words, in the absence of information on the timing of verification, we might mistakenly attribute an uptick in engagement to the Musk acquisition rather than the conferral of a blue tick. Given that both the Musk acquisition and blue tick verification are hypothesized to exert independent effects, it is important to be able to distinguish between them.

The list includes $\sim$138k accounts and consists of the account user name at the time of collection and their unique account ID. One of the developers involved in the original data release then merged these accounts with an existing project monitoring the activity of a list of "far-right and far-right-adjacent accounts" on Twitter.\footnote{See: \url{https://github.com/travisbrown/twitter-watch.}} This allows us to filter the 138k newly verified accounts by ranked centrality to these far-right networks. 

With these data, I take the top 1000 accounts based on rank centrality to far-right communication networks. With this information, I then use the \texttt{R} package \texttt{academictwitteR} \citep{barrie_academictwitter_2021} to retrieve all tweets posted by these accounts over the period May 17, 22 to November 23, 2022. In total, it was possible to gather data for 961 contentious accounts, giving $\sim$4.8m tweets.

We do not know the exact date of each account’s acquisition of blue-tick status. As a result, I use November 11 as my cut-off in the analysis. We are interested in any engagement boost that followed: 1) Musk’s acquisition of Twitter on October 27; 2) paid blue tick verification on or after November 11. 

I first use a panel setup where the unit of analysis is the user-day. The panel runs from May 17, 2022 to November 29, 2022. I use ordinary least squares (OLS) regression with standard errors clustered on the user. The estimating equation then takes the following functional form: 
\begin{equation}
\label{equation-baseline-fe}
\text{Retweets\,(logged)}_{i,t} = 
\beta\text{Musk\,Acquisition}_{i,t} +
\gamma\textbf{X}_{i,t} +
\alpha_i + \epsilon_{i,t}
\end{equation}
where $i$ represents individual contentious Twitter users, 
$t$ indexes our unit of observation (here: user-days), 
\textbf{X} is a control variable representing the total number of tweets posted by user on day $t$. The Musk acquisition is operationalized as a binary variable coded 1 if the date of the post is on or after October 27, and 0 otherwise. I plot all of the data for retweet and like dependent variables in Appendix Figures \ref{fig:alldatrts-ranked} and \ref{fig:alldatlikes-ranked}.

It is important to include the number of tweets posted by a given user on a certain day as this means we are able to estimate daily tweet engagement independent of activity. By introducing user-level fixed effects, we are recovering within-user estimates. In other words, we are able to recover the variation relative to what we might expect for a given user.

To determine the persistence of any effect, I alter the estimating equation to include a set of dummy variables for the sequence of three-day periods following the acquisition. In this model, the Musk acquisition is coded 1 on October 27, and 0 otherwise. We can conceptualize this model as a type of interrupted time-series design where the Musk acquisition binary represents the time of first treatment and the three-day time dummies a set of variables measuring time since treatment. In other words, the equation takes the following form: 
\begin{equation}
\label{equation-time-trends-fe}
\text{Retweets\,(logged)}_{i,t} = 
\beta\text{Musk\,Acquisition\,Day}_{i,t} +
\beta\text{Three-day\,Dummy}_{i,t} +
\gamma\textbf{X}_{i,t} +
\alpha_i + \epsilon_{i,t}
\end{equation}

\section{Results} \label{results}

Estimates are displayed in Table \ref{regtab}. We see that the acquisition of Twitter by Elon Musk is associated with significant increase in engagement on posts by contentious users engagement. This effect is net of any increased activity on the platform and is more pronounced for retweets than it is for likes. 

\begin{table}[htbp]
\caption{Effect of Musk acquisition on contentious user engagement}
\label{regtab}
\centering
\begin{tabular}[t]{lcc}
\toprule
  & Retweets & Likes\\
\midrule
Post-Musk acquisition & \num{0.536}*** & \num{0.128}***\\
 & (\num{0.027}) & (\num{0.020})\\
Tweets (logged sum) & \num{1.968}*** & \num{1.416}***\\
 & (\num{0.025}) & (\num{0.019})\\
\midrule
Observations & \num{187395} & \num{187395}\\
User fixed effect & X & X\\
\bottomrule
\multicolumn{3}{l}{\rule{0pt}{1em}Standard errors clustered by user}\\
\multicolumn{3}{l}{\rule{0pt}{1em}Outcomes are logged retweets and likes}\\
\multicolumn{3}{l}{\rule{0pt}{1em}* p $<$ 0.05, ** p $<$ 0.01, *** p $<$ 0.001}\\
\end{tabular}
\end{table}

Exponentiating these coefficients, we can say that the Musk acquisition saw a 70\% increase in user engagement in the form of retweets and a 14\% increase in likes.

\begin{figure}
    \includegraphics[width=14cm]{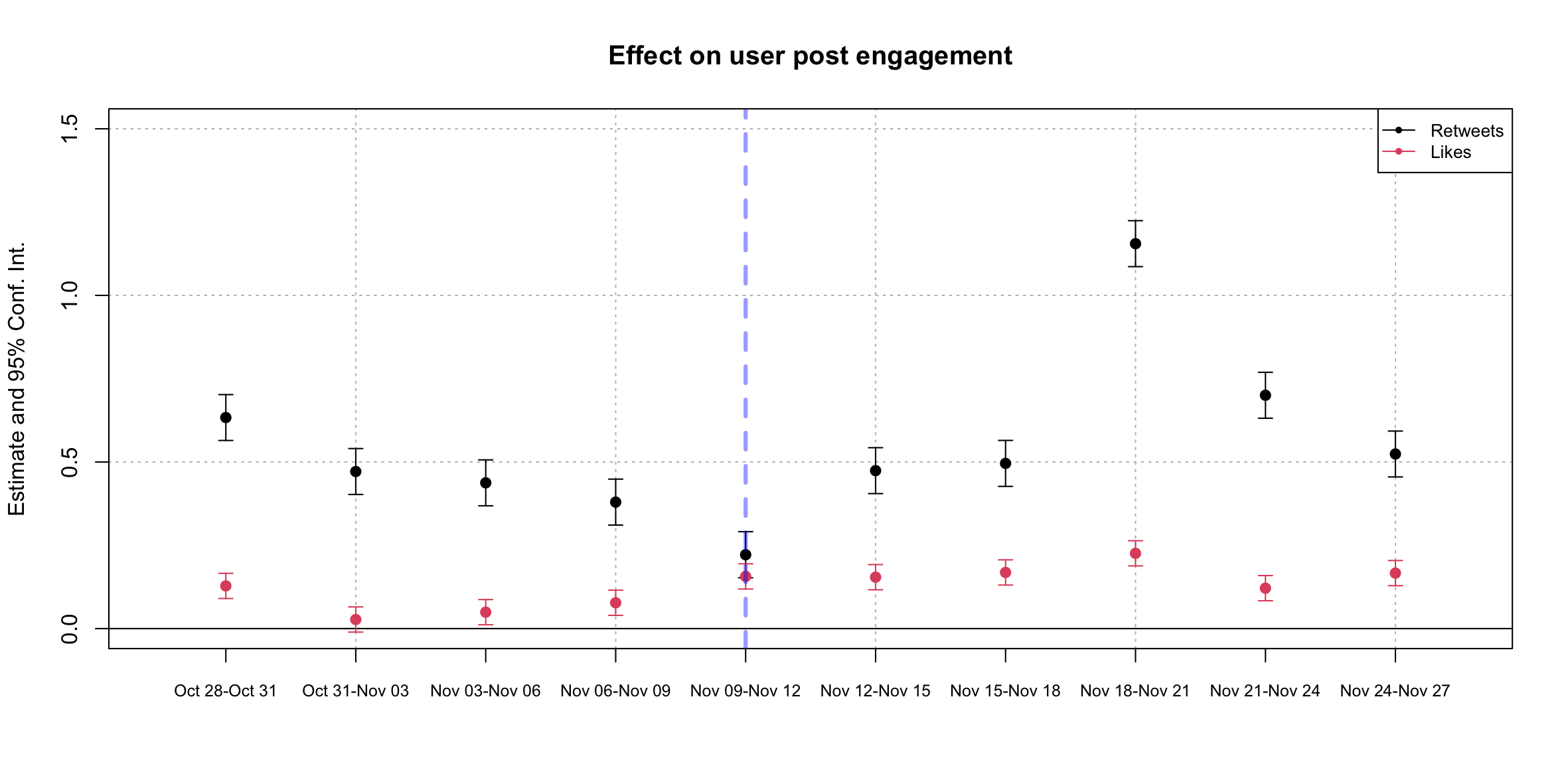}
    \caption{Three-day time dummy coefficients after Musk acquisition. Blue line represents approximate date of blue-tick verification.}
    \label{fig:efplot-ranked}
\end{figure}

Figure \ref{fig:efplot-ranked} displays coefficients for the three-day time period dummies over the month-long period following the Musk Acquisition. We see a sustained increase in post engagement, which dips around the time of verification applications. The subsequent uptick returns to the same levels of engagement as before the opening of blue tick verification. In sum, there is no obvious evidence of any immediate additional effect of blue tick verification on post engagement. 

Given we have no comparison group, one could argue that the effects we're observing result from a general increase in Twitter traffic that coincided with the Musk acquisition and changed verification policies. To test this conjecture, I also collect a sample of 1.2m tweets from 933 users who purchased blue tick verification but who are not counted in the reference list of contentious right-wing accounts. For these accounts, we see that the period following the Musk acquisition also saw an increase in post engagement but it was noticeably smaller: a 28\% increase in retweets and 4\% increase in likes. These results are displayed in Table \ref{regtab-verified} and three-day time dummy plots displayed in Appendix Figure \ref{fig:efplot-verified}.
\section{Discussion} \label{discussion}

Given the public pronouncement of the new Twitter owner about liberal bias and content policing, the acquisition of Twitter by Elon Musk in October, 2022, was heralded by some quarters as a victory for free speech.

In this article, I hypothesized that public signal of openness to previously moderated forms of speech was likely to boost engagement on accounts associated with contentious actors---particularly from the political right. In addition, I hypothesized that the conferral of blue tick verification to these accounts would provide an additional boost to engagement on content posted by these actors.

I find that in the period after Musk's acquisition of Twitter, contentious actors on the platform do see a sizeable boost in post engagement. I explain this as resulting from a simultaneous increase in users sympathetic to the views of these accounts, which resulted from Musk's public signals of intent to overhaul existing moderation policies. Conversely, I do not find evidence that blue tick verification was associated with any additional increase in post engagement.  

Another explanation for the results we observe is that there is some algorithmic bias baked into the Twitter algorithm. This would align with recent research that finds a modest right-wing bias in the Twitter amplification algorithm \citep{huszar_algorithmic_2022}. This would mean that any effect attributable to the Musk acquisition or blue-tick verification contains some additional algorithmic amplification. Unfortunately, without further experimental inquiry, it is not possible to disentangle observed effects from any algorithmic boost given to right-wing actors.

Finally, it is worth noting that the use of time dummies does not allow us to fully disentangle the effect of the Musk acquisition and blue-tick verification. This is because any signalling effect resulting from the Musk acquisition will also operate during and after the period users could purchase verification. Consistent across all estimates, however, is that the Musk acquisition and aftermath saw a sustained boost in engagement for content posted by contentious users---and this boost outstripped any concurrent increase for general users of the platform. 
\subsection{Acknowledgment}
I am grateful to Travis Brown for the public service of providing the original data used in this paper and for answering my queries. This preprint was created using the LaPreprint template (\url{https://github.com/roaldarbol/lapreprint}) by Mikkel Roald-Arb\o l
\clearpage
\printbibliography

\if@endfloat\clearpage\processdelayedfloats\clearpage\fi

\begin{appendix}

\section{Appendix}
\begin{center}
    \includegraphics[width=.75\linewidth]{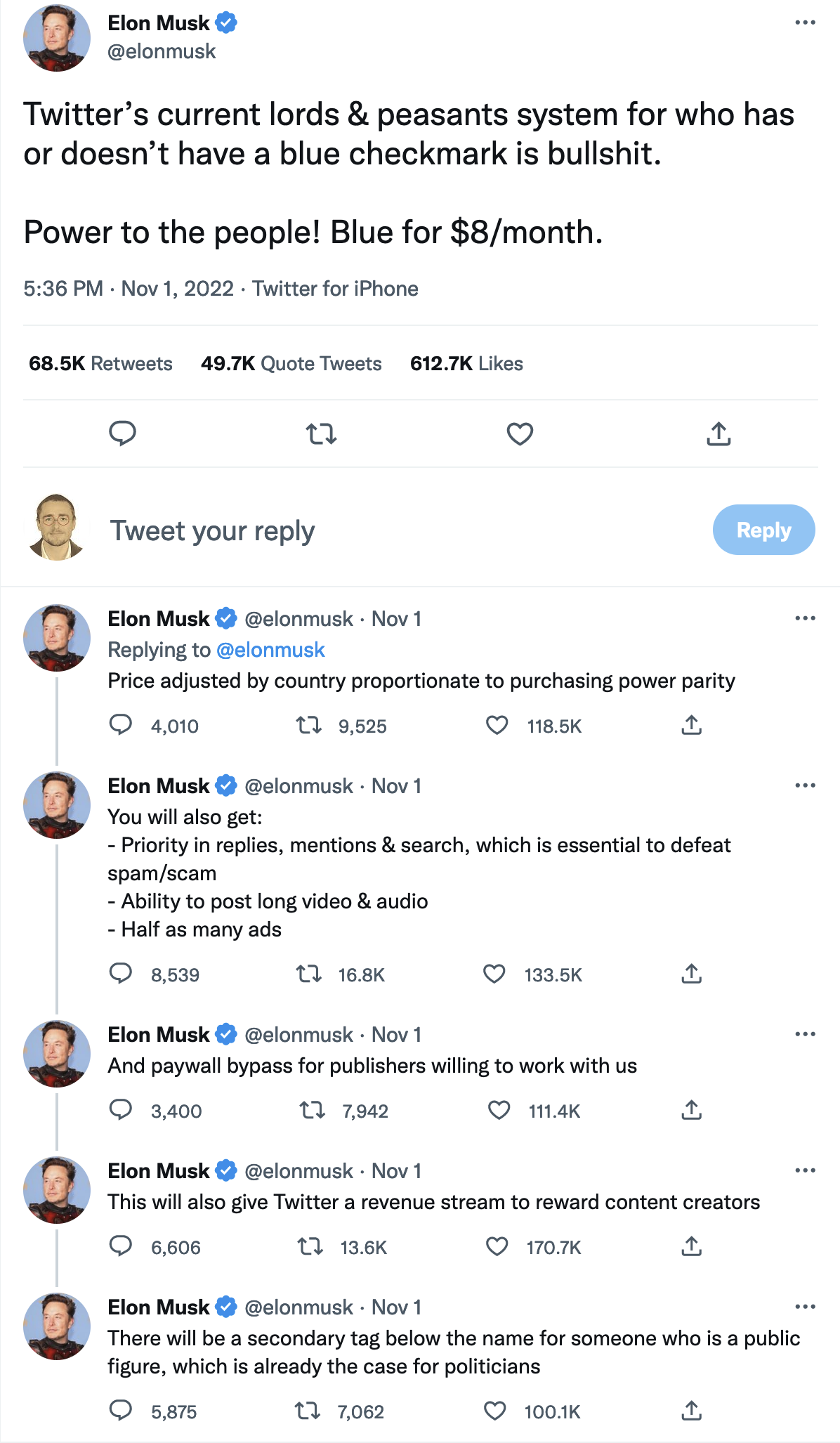}
    \captionof{figure}{Full tweet thread by Elon Musk posted on Nov 1, 2022}
    \label{fig:musktweetlong}
\end{center}

\begin{center}
    \includegraphics[width=.75\linewidth]{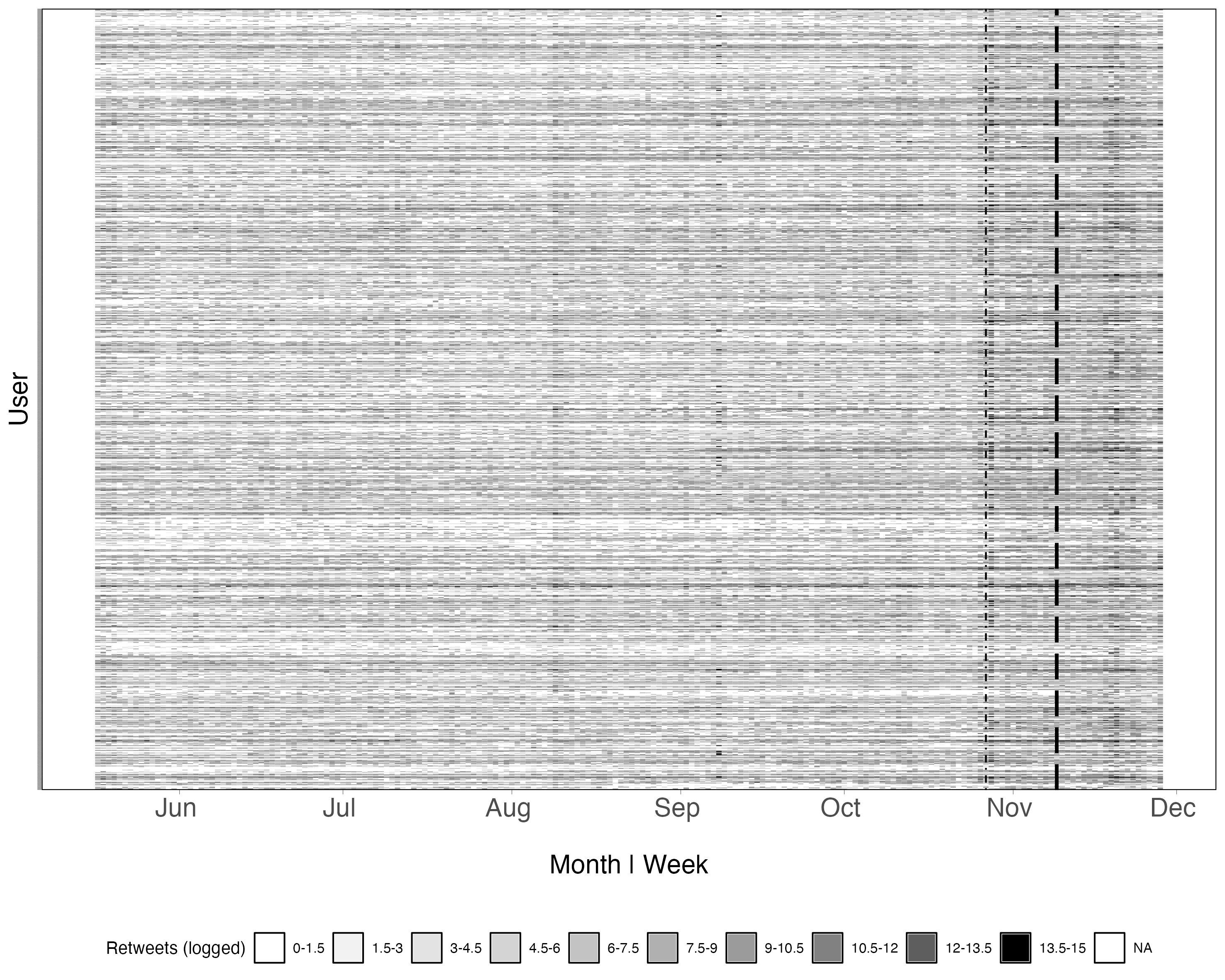}
    \captionof{figure}{User-day tile plot of retweets, May 17--Nov29, 2022 for contentious users}
    \label{fig:alldatrts-ranked}
\end{center}

\begin{center}
    \includegraphics[width=.75\linewidth]{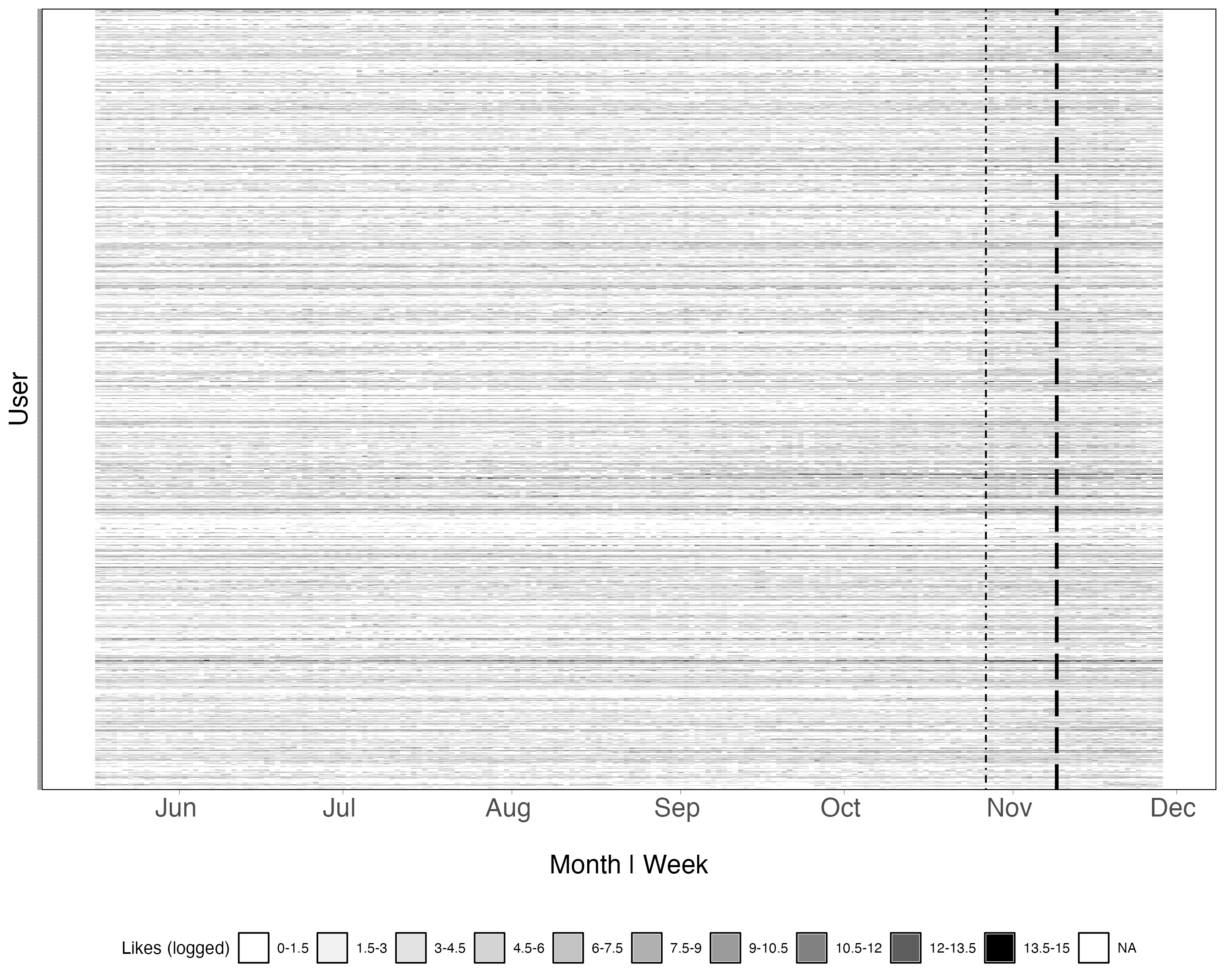}
    \captionof{figure}{User-day tile plot of likes, May 17--Nov29, 2022 for contentious users}
    \label{fig:alldatlikes-ranked}
\end{center}

\newpage
\begin{table}
\caption{Effect of Musk acquisition on general user engagement}
\label{regtab-verified}
\centering
\begin{tabular}[t]{lcc}
\toprule
  & Retweets & Likes\\
\midrule
Post-Musk acquisition & \num{0.246}*** & \num{0.036}***\\
 & (\num{0.025}) & (\num{0.010})\\
Tweets (logged sum) & \num{1.913}*** & \num{1.290}***\\
 & (\num{0.033}) & (\num{0.015})\\
\midrule
Observations & \num{181935} & \num{181935}\\
User fixed effect & X & X\\
\bottomrule
\multicolumn{3}{l}{\rule{0pt}{1em}Standard errors clustered by user}\\
\multicolumn{3}{l}{\rule{0pt}{1em}Outcomes are logged retweets and likes}\\
\multicolumn{3}{l}{\rule{0pt}{1em}* p $<$ 0.05, ** p $<$ 0.01, *** p $<$ 0.001}\\
\end{tabular}
\end{table}

\begin{center}
    \includegraphics[width=.75\linewidth]{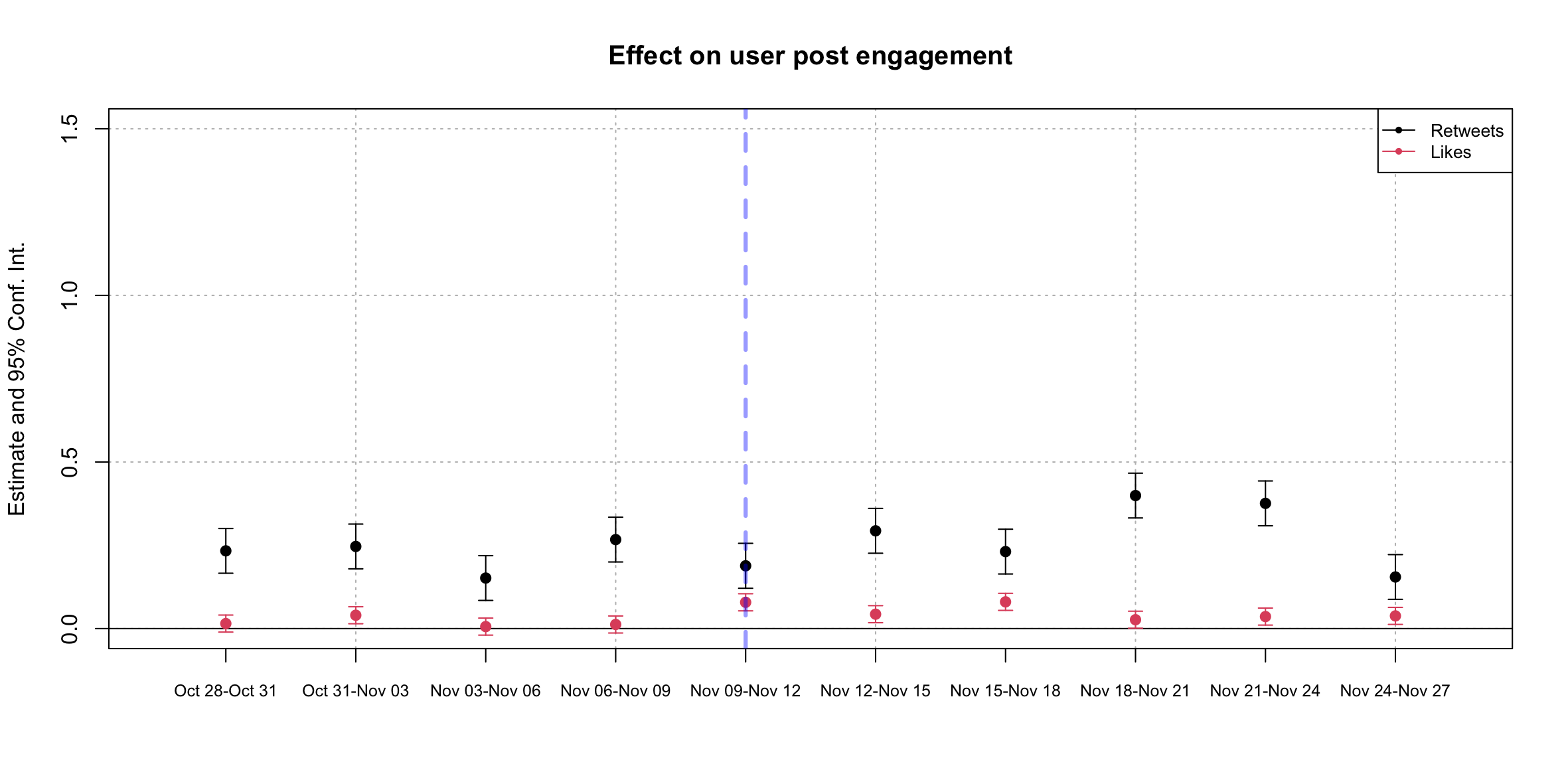}
    \captionof{figure}{Three-day time dummy coefficients after Musk acquisition. Blue line represents approximate date of blue-tick verification.}
    \label{fig:efplot-verified}
\end{center}

\end{appendix}


\end{document}